\documentstyle[prd,aps,floats]{revtex}

\input epsf

\begin{document}

\twocolumn[\hsize\textwidth\columnwidth\hsize\csname@twocolumnfalse\endcsname \draft 
\title{Vibrational Spectrum of the B=7 Skyrme Soliton} 
\author{W. K. Baskerville} 
\address{Department of Mathematical Sciences, University of Durham,
Science Laboratories, South Road, Durham, DH1 3LE, U.K. } 
\date{\today} \maketitle

\begin{abstract}
  The finite energy vibrational normal modes of the baryon number
  B=7 Skyrme soliton are computed. The structure of the spectrum
  obtained displays considerable similarity to those previously
  calculated for baryon numbers 2, 3 and 4. All modes expected from an
  approximate correspondence between skyrmions and BPS monopoles are
  found to be present. However, in contrast to earlier calculations,
  they do not all have energies below the pion mass. The remaining
  `breather-type' modes also conform to predictions, except that one
  predicted multiplet is not observed.
\end{abstract}

\vskip .2in 
]

\section{Introduction}
\label{sec:intro}

The normal mode spectra of multiskyrmions are believed to play a key
role in their quantisation \cite{uspub}. In the Skyrme model
\cite{Skyrme1}, nuclei appear as topological solitons in a classical,
non-linear field theory of $\pi$-mesons.  Since the model is not
renormalisable, a proper treatment as a quantum field theory is very
difficult. The usual approach to quantisation is based on the
suggestion of Manton, that the low energy dynamics of $n$ skyrmions
may be approximated by motion on some finite dimensional manifold. The
simplest possibility is the moduli space generated by the zero modes
of the minimal energy solution. For B=1, collective coordinate
quantisation of (coupled) spin and isospin gives a description of
nucleons and the $\Delta$ resonance in qualitative agreement with
experiment\cite{ANW,AN}. However, such a quantisation includes effects
of order $\hbar^{2}$ while ignoring effects of order $\hbar$. Also, it
is known that for real nuclei, nucleons in the ground state have
relatively large kinetic energies. Hence, the number of degrees of
freedom quantised must {\em at least\/} equal the number of nucleon
coordinates. Widely separated skyrmions can be labelled by their
positions and relative isospin orientations, so a minimum of $6B$
degrees of freedom are required to describe a Skyrme configuration of
baryon number B. However, the minimum energy always occurs when the
skyrmions coalesce into a single large soliton \cite{BC,BatSut},
with a maximum of 9 zero modes: translation, spin and
isospin. Clearly, in the neighbourhood of such a configuration, the
moduli space approximation must be extended to include low-lying
vibrational modes.

The first step in this direction was taken by Leese et al \cite{LMS},
who quantised a ten-dimensional manifold along the B=2 attractive
channel, using an instanton approximation for the field
configurations. While still not incorporating quite enough degrees of
freedom, their results were significantly better than had previously
been obtained from a simple quantisation of the eight zero
modes. Walet \cite{Walet} later extended this treatment, again
employing the instanton approximation, to estimate the full normal
mode spectra for B=2 and B=3. The vibrational spectra for B=2, 3 and 4
have now been directly computed by Barnes et al \cite{uspub,usb3}, who
were also able to classify the modes according to the representations
they form of the symmetry groups of the static solutions. A remarkable
pattern emerged. The lower frequency modes correspond to known
attractive channel scatterings; it was also observed that these modes
fall into representations exactly corresponding to those for small
zero mode deformations of BPS monopoles, for monopole charges equal to
the baryon numbers of the multiskyrmions. This strongly suggests an
underlying link between the two theories, and in fact Houghton et al
\cite{HMS} have demonstrated a connection through rational maps. They
speculate that the correspondence between BPS monopoles and skyrmions
should hold for any baryon number (monopole charge), and have used
this assumption to predict the lowest twenty one modes for the B=7
skyrmion.

The remaining, higher frequency parts of the B=2, 3 and 4 spectra
consist of `breathing'-type modes. The first mode above the pion mass
is in all cases a trivial breather, a simple size fluctuation of the
multiskyrmion. Above this lie more complicated breathing modes, where
one part of the soliton expands, while another part is
compressed. Baskerville and Michaels \cite{BM} recently proposed a
simple geometric explanation of these modes, based on the movement of
`branch lines': lines of (approximately) zero baryon density radiating
out from the centre of the minimal energy solutions. This ansatz
explains all the higher frequency normal modes observed for B=2, 3 and
4, although it also predicts an extra triplet for B=4 which was not
observed numerically. It is possible that this mode, if it exists,
might have a rather high frequency, which could explain why it was not
seen. Baskerville and Michaels also published a complete set of
predictions for the vibrational spectra of B=5, 6 and 7
multiskyrmions. In general, their geometrical ansatz predicts $4{\rm
B}-7$ higher breathing modes, in addition to a trivial breather, for a
minimal energy Skyrme solution of baryon number B, if ${\rm B} >
2$. The monopole analogy of Houghton et al predicts a further $4{\rm
B}-7$ lower frequency, scattering-type modes. Added to the nine zero
modes (for ${\rm B} > 2$), this gives a total of $8{\rm B} - 4$
degrees of freedom. This is rather more than the 6B zero modes of B
widely separated skyrmions, throwing open the question as to how many
degrees of freedom are really relevant in the vicinity of the minimal
energy solution, and therefore how many should be quantised. It is
interesting to note that instanton generated approximations to Skyrme
field configurations naturally have $8{\rm B}-1$ degrees of
freedom. Houghton \cite{Houghton} recently computed the vibrational
spectra of the instanton-generated 3-skyrmion, and obtained the same
modes as predicted by a combination of the monopole and geometrical
methods, plus an additional triplet, with axial vector symmetry.

In this article, the vibrational normal modes of the B=7 skyrmion
are computed, using the same method as in \cite{uspub,usb3}. Their
symmetries are then classified according to the symmetry group of the
static soliton, in this case the icosehedral group, the minimal energy
solution having the shape of a perfect dodecahedron \cite{BatSut}. The
B=7 spectrum is of particular interest since a full set of
predictions already exists for the normal modes. It thus provides a
good test of both the monopole analogy and the geometrical ansatz for
higher modes. In fact, all of the expected vibrations are observed,
except for one five-fold degenerate multiplet. The B=7 vibrational
spectrum may also be important for another reason. Irwin \cite{Irwin}
recently performed a zero mode quantisation of minimal energy
skyrmions for baryon numbers four to nine and seventeen. For baryon
numbers four, six and eight he obtained ground states in agreement
with experimental observations. However, for baryon numbers five,
seven, nine and seventeen, the calculated ground states have the wrong
spins. For the specific case of B=7, it may be possible to resolve
the problem through the inclusion of vibrational modes. The
experimentally observed ground state is an isodoublet with spin
$\frac{3}{2}$, whereas the lowest allowed state in the skyrmion
quantisation is an isodoublet with spin $\frac{7}{2}$. However, it is
possible that a spin $\frac{3}{2}$ rotational state could be combined
with a vibrational state to give an allowed state. If the vibrational
energy involved was small enough, the resultant state might have lower
energy than the spin $\frac{7}{2}$ isodoublet. Obviously, knowledge of
the exact frequencies and symmetries of the low-lying vibrational
states is a prerequisite for progress in this direction.

\section{Computation and Interpretation of the Spectrum}
\label{sec:method}

The normal mode spectrum is computed using the same method and
computer codes developed for the B=2, 3 and 4 calculations
\cite{uspub,usb3}. Skyrme fields, being SU(2)-valued, lie on a
3-sphere, and can thus be represented by a scalar field $\phi \in
\Re^{4}$, with the constraint $\phi^{a} \phi^{a} = 1$. The Skyrme
Lagrangian density can then be written, in terms of the usual
dimensionless Skyrme units \cite{skyrme2}, as
\begin{eqnarray}
  {\cal L}  & = & \frac{1}{2}\, \partial_{\mu}\phi \, \partial^{\mu}\! \phi 
  \ + \ \omega_{\pi}^{2} \phi^{1} \ + \ \lambda(\phi \cdot \phi - 1) \\
  \label{Ldef}
  \ & + & \frac{1}{4} \{(\partial_{\mu} \phi \cdot \partial_{\nu} \phi )
  (\partial^{\mu}\! \phi \cdot \partial^{\nu}\! \phi)\! -\! 
  (\partial_{\mu} \phi \cdot \partial^{\mu}\! \phi)
  (\partial_{\nu} \phi \cdot \partial^{\nu}\!\phi) \} \nonumber
\end{eqnarray}
where $\lambda$ is a Lagrange multiplier field, and $\omega_{\pi} =
\frac{2m_{\pi}}{F_{\pi}e}$ is a dimensionaless constant, the
oscillation frequency of the homogeneous pion field. As in
\cite{uspub,usb3}, the `standard' value \cite{AN} $\omega_{\pi} =
.526$ has been used.

The Skyrme equations of motion are derived by discretising the action
$S = \int \!d^{4}\!x \, {\cal L}$ on a finite lattice, then varying
with respect to each field component $\phi^{a}(t,i,j,k)$, at each
point in the lattice. This leads in general to a coupled system of
equations, due to the mixed time and space derivatives in the quartic
Skyrme term. The relevant, ``kinetic'' part of the Lagrangian density
can be written
\begin{equation}
  {\cal L}_{\mbox \small{\rm KIN}} = \frac{1}{2} \stackrel{\mbox{{\huge
  .}}}{\phi}^{a} K^{ab}(\partial_{i}\phi) \stackrel{\mbox{{\huge
  .}}}{\phi}^{b}
  \label{Lkin}
\end{equation}
where $K^{ab}= \delta^{ab} (1+(\partial_{i}\phi)^2) -
\partial_{i}\phi^a \partial_{i}\phi^b$ acts as a spatially dependent
inertia tensor. It is the time dependence of $K^{ab}$ which leads to
the coupling, with all its inherent numerical difficulties. However,
for very small perturbations about the static solution, $\phi(t,{\bf
x}) = \phi_{\mbox \small {\rm st}}({\bf x}) + \varepsilon(t,{\bf x})$,
the time dependent parts of $K^{ab}$ give rise to terms of order at
least $\varepsilon^{3}$. If $\varepsilon \ll 1$, such terms can be
neglected to a good approximation, and the matrix $K^{ab}$ assigned
its value at the static classical solution. The discretised equations
of motion then reduce to the form
\begin{equation}
  \phi^{a}(t+1,i,j,k) = R^{a}(t,i,j,k) + \tilde{\lambda} \phi^{a}(t,i,j,k)
  \label{eofm}
\end{equation}
where $\tilde{\lambda}$ is a constant related to the Lagrange
multiplier, and $R^{a}$ is a function of the fields at the current and
previous timesteps only. Invoking the SU(2) constraint $\phi^{a}_{t+1}
\phi^{a}_{t+1} =1$ yields a quadratic equation for $\tilde{\lambda}$
\begin{equation}
  \tilde{\lambda}^{2} + 2 \tilde{\lambda} R^{a} \phi^{a} + 
  R^{a} R^{a} - 1 = 0  ,
  \label{lambdadef}
\end{equation}
the positive square root of which is the desired solution (the
definitions of $\tilde{\lambda}$ and $R^{a}$ can be chosen so as to
ensure this: see \cite{PST} for details). Substituting
$\tilde{\lambda}$ back into Eq.~(\ref{eofm}) then gives a
deterministic time evolution for the Skyrme fields.

The vibrational spectrum can be directly computed using this
algorithm. The static solution is found by numerical relaxation, using
fields derived from the rational map ansatz as a starting point
\cite{HMS}. A small random perturbation is then applied, after which
the fields are evolved forward for a long time at constant energy. The
evolving fields take the form
\begin{equation}
  \phi(t,{\bf x}) = \phi_{\mbox \small{\rm st}}({\bf x})
  + \sum_{\mbox \small{\rm modes}} \epsilon_{n} \delta_{n}({\bf x})
  \cos(\omega_{n} t) + {\bf O}(\epsilon^{2}) ,
  \label{fieldev}
\end{equation}
where the functions $\delta_{n}({\bf x}) \in \Re^{4}$, obeying
$\delta_{n}({\bf x}) \cdot \phi_{\rm st}({\bf x}) = 0$ are the normal
modes, each excited with amplitude $\epsilon_{n}$. The normal mode
frequencies $\omega_{n}$ are obtained by Fourier analysis of the field
$\phi(t,{\bf x})$, with respect to time, at any point in the box.

Once the frequencies have been identified, the time evolution process
is repeated. Maps of the normal modes $\delta_{n}({\bf x})$ can be
constructed by performing discrete Fourier sums on each component of
the evolving field. The space of perturbations has a very useful inner
product
\begin{equation}
  \langle \delta_{1} | \delta_{2} \rangle = \int_{\mbox \small{\rm box}}
  \delta^{a}_{1}({\bf x}) K^{ab}({\bf x}) \, \delta^{b}_{2}({\bf x})
  \:{\rm d}^{3}{\bf x} ,
  \label{innerprod}
\end{equation}
the matrix $K^{ab}({\bf x})$ being included in this definition to
ensure orthogonality between modes of different frequency: $\langle
\delta_{1} | \delta_{2} \rangle = 0$ if $\omega_{1} \ne \omega_{2}$.
In general, the normal modes fall into degenerate multiplets at
discrete frequencies. Each such multiplet forms a matrix
representation of the symmetry group of the static soliton, via the
inner product $\langle \delta_{i} | S | \delta_{j} \rangle$. The
symmetry operations $S$ in general consist of a physical operation,
for example a rotation or reflection, which leaves the energy and
baryon densities of the static soliton invariant, plus an isospin
transformation of the pion fields ($\phi^{2}$, $\phi^{3}$,
$\phi^{4}$), which ``undoes'' the effect of the physical operation on
the static fields. Modes of the same frequency always form an
irreducible representation (irrep) of the symmetry group, which places
strong constraints on the allowed degeneracies. In the case of the B=7
multiskyrmion, the symmetry group of the static soliton is the
icosehedral group, which has irreducible representaions of degeneracy
one, three, four and five only.

Determining the degeneracy and symmetry of each multiplet is in
principle straightforward. For degenerate modes, each random initial
perturbation will produce a different linear combination $\sum_{i}
\epsilon_{i} \delta_{i}({\bf x})$, when one projects out the relevant
frequency. Applying the symmetries of the static soliton will yield
yet other, different combinations. The degeneracy of a given frequency
can be found by computing the rank of the matrix of inner products
between different linear combinations of the degenerate modes. It is
also possible to project any given linear combination out of all the
others. In this way, a library of orthonormal modes can be built
up. Once a full orthonormal basis has been obtained for a given
frequency, it is easy to calculate its symmetry. Irreducible
representations are uniquely labelled by the characters (traces
$\sum_{i}\langle \delta_{i} | S | \delta_{i} \rangle$) of certain
symmetry operations $S$.

Unfortunately, several difficulties arise in this procedure for the
particular case of B=7. The main problem is that the simulations are
carried out on a finite, cubic lattice, which breaks the dodecahedral
symmetry of the static solution. The most immediate consequence of
this is that some multiplets may be artificially split, for example a
five-fold degenerate multiplet into a doublet and triplet of close,
but not identical, frequency. The breaking may also cause certain
modes to be generated more easily than others, if they fit better onto
the finite grid. Thus different initial conditions do not really
produce randomly different linear combinations of degenerate
modes. This can effectively `hide' some of the degeneracy, since the
method outlined above depends on finding enough {\em independent\/}
linear combinations. A closely related technical difficulty is that
not all of the symmetry operations of the icosehedral group can be
exactly realised on a cubic lattice. If the cubic lattice does not map
onto itself under some physical operation, then interpolation must be
used to find the new field values. This can be done accurately enough
to compute the character of such a symmetry, given an orthonormal
basis; but is not suitable for separating out such a basis in the
first place. Since only 24 (out of 120) physical symmetries can be
performed exactly on the cubic lattice, this is a severe
limitation. The last (but certainly not least) problem is
contamination between modes of different frequency. Forty-three
vibrational modes are predicted for B=7, lying in at least eleven
different multiplets. These are all expected to occur within a fairly
small range of frequencies. The finite time scale of the simulation
leads inevitably to finite-width peaks in the power
spectrum. Considerable overlap between different multiplets is
therefore possible, and does in fact occur, as can immediately be seen
from Figure~().

The resolution of these difficulties requires a closer examination of
the icosehedral symmetry group $I_{h}$, as well as the particular
alignment of the dodecahedral static solution on the cubic
lattice. The character table for $I_{h}$ is given in
Table~\ref{tab:char}. The physical symmetries can be divided into ten
`conjugacy classes' of like operations. The identity $\openone$ and
inversion $i$ both form classes on their own. There are six five-fold
rotational axes in a dodecahedron, running through the centres of
opposite faces.  Rotations by $\frac{2\pi}{5}$ about these, in either
direction, form one conjugacy class (with 12 elements), labelled
$C_{5}$, while rotations by $\frac{4\pi}{5}$ form another,
$C_{5}^{2}$, also with 12 elements. There are ten three-fold axes,
through pairs of opposite corners of the dodecahedron. The 20
$\frac{2\pi}{3}$ rotations associated with these form another
conjugacy class, $C_{3}$. The final rotational symmetry is two-fold,
about axes which bisect pairs of opposite edges of the
dodecahedron. This gives 15 rotations by $\pi$ (there are 30 edges,
since each edge is shared by two of the twelve pentagonal faces),
forming the conjugacy class $C_{2}$. The remaining symmetry operations
are all combinations of inversion with one of the rotations already
mentioned, and are denoted $iC_{n}$, for rotation $C_{n}$. (Note that
$iC_{2}$ gives pure reflections, in planes which contain two opposite
edges, and bisect four pentagonal faces.) The characters for these
classes are exactly the same as for the equivalent rotations without
inversion, except that negative parity representations pick up a minus
sign. This is because the full icosehedral group $I_{h}$ is just the
group of rotations of an icosehedron (dodecahedron) $I$, extended by
inversion: $I_{h} = I \times Z_{2}$. The irreducible representations
are labelled by their degeneracies and parities, where this is
sufficient to distinguish them (eg.\ $5^{-}$ is a five-fold degenerate
multiplet with negative parity). The only exceptions are the four
three-dimensional irreps, which are labelled $F_{1}$ and $F_{2}$,
again with superscripts denoting parity. With this notation, a vector
has symmetry $F_{1}^{-}$, while an axial vector transforms as
$F_{2}^{+}$.
\begin{table}[b]
  \begin{center} \leavevmode 
  \begin{tabular}{|c|c|c|c|c|c|c|c|c|c|c|} \hline 
    \ & $\openone$ & 12$C_{5}$ & 12$C_{5}^{2}$ & 20$C_{3}$ & 15$C_{2}$ &
     $i$ & 12$iC_{5}$ & 12$iC_{5}^{2}$ & 20$iC_{3}$ & 15$iC_{2}$ \\ \hline 
    $1^{+}$ & 1 & 1 & 1 & 1 & 1 & 1 & 1 & 1 & 1 & 1 \\ 
    $F_{1}^{+}$ & 3 & $\tau$ & $1-\tau$ & 0 & -1 &
     3 & $\tau$ & $1-\tau$ & 0 & -1 \\ 
    $F_{2}^{+}$ & 3 & $1-\tau$ & $\tau$ & 0 &
     -1 & 3 & $1-\tau$ & $\tau$ & 0 & -1 \\ 
    $4^{+}$ & 4 & -1 & -1 & 1 & 0 & 4 & -1 & -1 & 1 & 0 \\ 
    $5^{+}$ & 5 & 0 & 0 & -1 & 1 & 5 & 0 & 0 & -1 & 1 \\ \hline 
    $1^{-}$ & 1 & 1 & 1 & 1 & 1 & -1 & -1 & -1 & -1 & -1 \\ 
    $F_{1}^{-}$ & 3 & $\tau$ & $1-\tau$ & 0 & -1 & 
     -3 & $-\tau$ & $\tau -1$ & 0 & 1 \\ 
    $F_{2}^{-}$ & 3 & $1-\tau$ & $\tau$ & 0 & -1 & 
     -3 & $\tau -1$ & $-\tau$ & 0 & 1 \\ 
    $4^{-}$ & 4 & -1 & -1 & 1 & 0 & -4 & 1 & 1 & -1 & 0 \\ 
    $5^{-}$ & 5 & 0 & 0 & -1 & 1 & -5 & 0 & 0 & 1 & -1 \\ \hline 
  \end{tabular} 
  \end{center}
  \caption{Character table for $I_{h}$. $\tau = \frac{\sqrt{5} +1}{2}$.}
  \label{tab:char}
\end{table}

\begin{figure}[tb]
  \begin{center} 
    \leavevmode 
    {\hbox {\epsfxsize = 6cm \epsffile{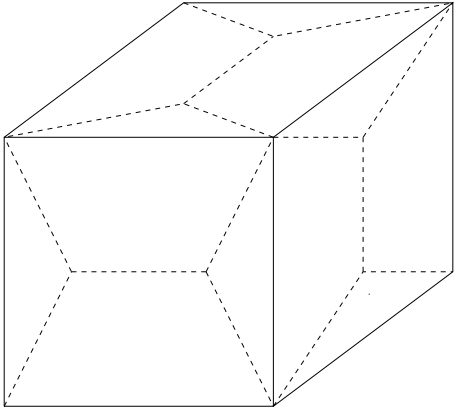}}} 
  \end{center} 
  \caption{Schematic diagram of one cube within a dodecahedron. The
    edges of the dodecahedron are drawn as dotted lines on the faces
    of the cube. In reality, they would project slightly, and the
    edges of the cube would lie slightly behind the pentagonal faces.}
  \label{fig:cube}
\end{figure}

Next, we need to examine the alignment of the dodecahedral soliton on
the cubic grid. A dodecahedron contains within itself five `cubes';
that is, there are five sets of eight vertices, each of which form the
corners of a cube. Each of the 20 vertices of the dodecahedron is a
member of two such cubes. In the B=7 static soliton used in
simulations, one of these cubes is aligned with the axes of the cubic
lattice. See Figure~(\ref{fig:cube}) for a schematic diagram. 
It should be obvious, looking at this Figure, which symmetry
operations map this cube to itself (and hence preserve the grid
structure): the identity, inversion, three $C_{2}$ rotations (about
the Cartesian axes), three reflections, eight $C_{3}$ rotations (about
the body diagonals of the cube), and the same eight rotations combined
with inversion. The vital question is how to use these symmetries to
produce orthonormal bases for all possible multiplets. It is very
useful, at this stage, to keep in mind some sort of physical model for
representations of various degeneracies. For example, the Cartesian
axes will obviously transform as a triplet (in fact
$F_{1}^{-}$). Given a general mixture of all three (a vector in a
random direction), the symmetry operations $(1\!\!1 + C_{x})$,
$(1\!\!1 + C_{y})$ and $(1\!\!1 + C_{z})$, or equivalently $(C_{y} +
C_{z})$, $(C_{x} + C_{z})$ and $(C_{x} + C_{y})$, will project out the
$x$, $y$ and $z$ components respectively, where $C_{x}$ is a
$180^{\circ}$ rotation about the $x$-axis, etc. Degeneracies four and
five are a little trickier. The five cubes mentioned above form a
representation, since all symmetries of the dodecahedron will rotate
them among themselves. This representation is not irreducible, since
the sum or symmetric combination of all five transforms trivially (as
$1^{+}$), ie.\ this combination is invariant under all physical
symmetries of the dodecahedron. However, the four remaining
antisymmetric configurations, orthogonal to the symmetric combination
and to each other, do give an irreducible $4^{+}$
representation. Labelling the five cubes $\{ |1 \rangle \cdots |5
\rangle \}$, a suitable orthonormal basis for the degeneracy 4 irrep
can be written
\begin{mathletters}
  \label{4basis}
\begin{equation}
  |A \rangle = \frac{1}{2}\, (|1 \rangle\! +\! |2 \rangle\! -\! 
   |3 \rangle\! -\! |4 \rangle )
\end{equation}
\begin{equation}
  |B \rangle = \frac{1}{2}\, (|1 \rangle\! +\! |3 \rangle\! -\! 
   |2 \rangle\! -\! |4 \rangle )
\end{equation}
\begin{equation}
  |C \rangle = \frac{1}{2}\, (|1 \rangle\! +\! |4 \rangle\! -\! 
   |2 \rangle\! -\! |3 \rangle )
\end{equation}
\begin{equation}
  |D \rangle = \frac{1}{2 \sqrt{5}}\, (4 |5 \rangle\! -\! |1 \rangle\! -\! 
   |2 \rangle\! -\! |3 \rangle\! -\! |4 \rangle ). 
\end{equation}
\end{mathletters}
If the cube which is aligned with the lattice is taken to be $|5
\rangle$, then the $C_{2}$ rotations about the Cartesian axes act as
follows:
\begin{mathletters}
  \label{4rotns}
\begin{equation}
C_{x}: (A,B,C,D) \longrightarrow (A,-B,-C,D)
\end{equation}
\begin{equation}
C_{y}: (A,B,C,D) \longrightarrow (-A,B,-C,D)
\end{equation}
\begin{equation}
C_{z}: (A,B,C,D) \longrightarrow (-A,-B,C,D) .
\end{equation}
\end{mathletters}
Using these results, it is quite simple to construct projection
operators for the orthonormal basis
\begin{mathletters}
  \label{4projectors}
\begin{equation}
  P_{A} = \openone + C_{x} - C_{y} - C_{z} \nonumber \\
\end{equation}
\begin{equation}
  P_{B} = \openone - C_{x} + C_{y} - C_{z} \nonumber \\
\end{equation}
\begin{equation}
  P_{C} = \openone - C_{x} - C_{y} + C_{z} \nonumber \\
\end{equation}
\begin{equation}
  P_{D} = \openone + C_{x} + C_{y} + C_{z}.
\end{equation}
\end{mathletters}
It should be noted at this stage that $P_{A}$, $P_{B}$ and $P_{C}$
also act as projectors for the triplet formed by the Cartesian
axes. What now for a five-fold basis? There are six $C_{5}$ rotational
axes, giving a 6-dimensional representation which decomposes in a
similar way to the 5 cubes: again the symmetric combination transforms
trivially as $1^{+}$, while five orthogonal antisymmetric combinations
form $5^{+}$. Labelling the six $C_{5}$ axes $\{ |1 \rangle \cdots |6
\rangle \}$, a possible basis for the latter is
\begin{mathletters}
  \label{5basis}
\begin{equation}
  |a\rangle = \frac{1}{\sqrt{2}}\,(|1\rangle\! -\! |2\rangle), \ \ \ \ \ 
  |b\rangle = \frac{1}{\sqrt{2}}\,(|3\rangle\! -\! |4\rangle ),
\end{equation}
\begin{equation}
  |c\rangle = \frac{1}{\sqrt{2}}\,(|5\rangle\! -\! |6\rangle ), \ \ \ 
  |d\rangle = \frac{1}{2}\, (|1\rangle\! +\! |2\rangle\! -\! 
  |3\rangle\! -\! |4\rangle ), 
\end{equation}
\begin{equation}
  |e\rangle = \frac{1}{2 \sqrt{3}}\, (2 |5\rangle\! +\! 2|6\rangle\! -\! 
   |1\rangle\! -\! |2 \rangle\! -\! |3 \rangle\! -\! |4 \rangle ).  
\end{equation}
\end{mathletters}
The action of $C_{x}$, $C_{y}$ and $C_{z}$ on these states is as
follows
\begin{mathletters}
  \label{5rotns}
\begin{equation}
C_{x}: (a,b,c,d,e) \longrightarrow (a,-b,-c,d,e) \nonumber \\
\end{equation}
\begin{equation}
C_{y}: (a,b,c,d,e) \longrightarrow (-a,b,-c,d,e) \nonumber \\
\end{equation}
\begin{equation}
C_{z}: (a,b,c,d,e) \longrightarrow (-a,-b,c,d,e) .
\end{equation}
\end{mathletters}
It is then apparent that the projection operators $P_{A}$, $P_{B}$ and
$P_{C}$ defined in Eq.~(\ref{4projectors}) also act as projectors for
states $|a\rangle$, $|b\rangle$ and $|c\rangle$ respectively, while
$P_{D}$ will project out some combination of states $|d\rangle$ and
$|e\rangle$. It may seem an incredible coincidence that the same
symmetries project out the basis states for multiplets of degeneracy
three, four and five, but in fact this is deeply related to the way in
which the cubic lattice breaks the icosehedral symmetry. Furthermore,
these same projectors work for {\em all} degeneracy three, four and
five multiplets, not just $F_{1}^{-}$, $4^{+}$ and $5^{+}$, from which
they were derived. This last is not obvious, in fact is perhaps
counter-intuitive, but nevertheless has been found to work in
practice.

The projectors defined in Eq.~(\ref{4projectors}) are the main tool
required to separate out the various multiplets, when contamination
occurs between them. States of different parity can easily be
separated, using the projectors
\begin{equation}
  P_{+} = \openone + i, \ \ \ \ \ \ \ \ \ \ P_{-} = \openone - i .
  \label{parity}
\end{equation}
However, contamination between degenerate multiplets of the same
parity is more awkward. The coincidence between the projection
operators means that mode $|A\rangle$ of a $4^{+}$ cannot be
distinguished from mode $|a\rangle$ of a $5^{+}$ for example, or from
the ``x'' component of $F_{1}^{+}$ or $F_{2}^{+}$. However, it {\em
is} possible to tell whether the three modes projected out by $P_{A}$,
$P_{B}$ and $P_{C}$ are derived purely from a single multiplet, or
represent some mixture, by considering their behaviour under $C_{3}$
rotations. A $120^{\circ}$ degree rotation in physical space causes a
cyclic permutation of the three projections. This is also a
$120^{\circ}$ degree rotation in the `phase space' of the three modes,
and since the modes may be aligned or anti-aligned with the Cartesian
axes of this space, two of the modes may pick up minus signs. This
occurs randomly, and the chance of it happening the same way for two
different multiplets which are mixed together is only
$\frac{1}{64}$. Hence if the overlap $\langle \delta | P_{A} | \delta
\rangle$ is {\em exactly\/} equal to $\pm \langle \delta | C_{3} P_{B}
| \delta \rangle$, we are almost certainly dealing with a pure
multiplet. If, furthermore, $\langle \delta | P_{D} | \delta \rangle =
0$, then $|\delta\rangle$ represents a pure triplet. Otherwise, the
mode must have degeneracy four or five. The latter possibilities can
be distinguished by the behaviour of $P_{D} |\delta \rangle$ under
$C_{3}$ rotations. If the multiplet has degeneracy four, then $\langle
\delta | P_{D}^{\dagger} C_{3} P_{D} | \delta \rangle = 1$, while for
a degeneracy 5 multiplet $\langle \delta | P_{D}^{\dagger} C_{3} P_{D}
| \delta \rangle = -\frac{1}{2}$. For degeneracy 5, the operator
$(C_{3} - C_{3}^{-1}) P_{D}$ ( where $C_{3}$ and $C_{3}^{-1}$ are
$120^{\circ}$ rotations in opposite directions about the same axis)
projects out a linear combination orthogonal to that projected out by
$P_{D}$, thus completing the orthonormal basis.

The overall procedure for interpretation of the spectrum is then as
follows. First of all, any contamination by zero modes (spin or linear
momentum) is projected out. A careful watch is kept during the
simulation to make sure that ``drift'' (ie.\ excitation of zero modes)
is kept within strict limits, to ensure the validity of the basic
assumption underlying the computer algorithm. Additionally, severe
damping is applied at the edges of the box for the first 1000
timesteps of the simulation, before any data collection begins. This
helps to decrease the proportion of radiation relative to real
vibrational modes, and also helps reduce the subsequent drift. Still
slight contamination from zero modes occurs, which must be removed
from the perturbations $\delta^{a}({\bf x})$. Next, the perturbations
are separated into positive and negative parity components. The
projectors $P_{A}$ to $P_{D}$ are applied to the resulting states. The
overlaps $\langle \delta | P_{i} |\delta\rangle$ and $\langle\delta|
C_{3} P_{i} |\delta\rangle$ are calculated, as is the matrix element
$\langle \delta | P_{D}^{\dagger} C_{3} P_{D} | \delta \rangle$. If a
pure multiplet is found, an orthonormal basis for it is
constructed. This can then be projected out of all other
frequencies. Hopefully this will remove enough contamination that at
least one other mode (of the same parity) will now be pure. The
procedure can then be repeated, until there are no pure modes left. As
the library of basis states for pure multiplets is built up, a check
is kept on how much of each frequency is accounted for. If all modes
are suitably normalised, then
\begin{equation}
  \sum_{i,n} \:\langle \delta | \omega_{n}^{(i)} \rangle 
  \langle \omega_{n}^{(i)}|\delta \rangle = 1
  \label{norm}
\end{equation}
for all original perturbations $|\delta\rangle$, where $\{
|\omega_{n}^{(i)}\rangle\}$ is the full set of orthonormal basis
states. It may happen that some multiplets are inextricably mixed, so
that it is not possible to extract a pure set of basis states for
them. This will apply mainly to the ``triplet'' states, as modes
$|D\rangle$ of $4^{\pm}$ and $\{|d\rangle, |e\rangle\}$ of $5^{\pm}$
can always be separated at least from each other. It is still possible
to tell how many ``triplets'' there are. Applying $P_{A}$, $P_{B}$ and
$P_{C}$ to any one of the remaining frequencies will give basis
vectors in the combined space of the remaining multiplets. This can be
projected out of the other frequencies, and the process repeated, as
for the pure modes, until all modes have been accounted for. In
practice, this means until all the sums in Eq.~(\ref{norm}) add to at
least 0.995. Thus it is always possible to fully classify the
symmetries of all vibrational modes in the spectrum, even if they
cannot all be projected out cleanly. The main difference is that
frequencies can be confidently assigned to the pure multiplets,
whereas mixtures will in general merely be confined within a certain
range of frequencies. Sometimes a tentative assignment can be made if
a pure $|D\rangle$ mode or $\{|d\rangle, |e\rangle\}$ doublet appears
strongly at a given frequency.

One last technical detail should be mentioned. 
As was explained above, the symmetry operations that will be applied
to the perturbations $\{|\delta\rangle \}$, are combinations of
physical symmetries and isospin transformations. The alignment of the
static solution on the cubic grid is as shown in
Figure~\ref{fig:cube}. It is also necessary to determine the exact
symmetry of the pion fields in the static solution, or at least, how
they transform under the specific symmetries needed to identify the
perturbations. From the foregoing discussion, it should be clear that
the symmetry operations needed to project out orthonormal bases for
all modes are inversion, the three $\pi$-rotations $C_{x}$, $C_{y}$
and $C_{z}$, plus one $\frac{2\pi}{3}$- rotation (any of the eight
which can be realised on the cubic lattice will do) and its
inverse. Additionally, we will need to calculate the character of one
$\frac{2\pi}{5}$ rotation, to distinguish between the triplet
representations $F_{1}$ and $F_{2}$. The following equations give the
full symmetry operations used, both the physical operations on the
Cartesian axes $(x,y,z)$, and the isospin transformations which {\em
restore} the original fields of the static solution.
\begin{mathletters}
  \label{pisymm}
\begin{eqnarray}
  i: \ & & (x,y,z) \longrightarrow (-x,-y,-z) \nonumber \\
  & & (\pi_{1},\pi_{2},\pi_{3}) \longrightarrow
      (-\pi_{1},-\pi_{2},-\pi_{3})
\end{eqnarray}
\begin{eqnarray}
  C_{x}: \ & & (x,y,z) \longrightarrow (x,-y,-z) \nonumber \\
  & & \left( \begin{array}{c}
        \pi_{1} \\ \pi_{2} \\ \pi_{3}
      \end{array} \right) \longrightarrow 
           \left( \begin{array}{ccc}
             \cos \theta & 0 & -\sin \theta \\
             0 & -1 & 0  \\
             -\sin \theta & 0 & -\cos\theta
           \end{array} \right)\!\!\!
                \left( \begin{array}{c}
                  \pi_{1} \\ \pi_{2} \\ \pi_{3}
                \end{array} \right)
\end{eqnarray}
\begin{eqnarray}
  C_{y}: \ & & (x,y,z) \longrightarrow (-x,y,-z) \nonumber \\
  & & (\pi_{1},\pi_{2},\pi_{3}) \longrightarrow (-\pi_{1},\pi_{2},-\pi_{3})
\end{eqnarray}
\begin{eqnarray}
  C_{z}: \ & & (x,y,z) \longrightarrow (-x,-y,z) \nonumber \\
  & & \left( \begin{array}{c}
        \pi_{1} \\ \pi_{2} \\ \pi_{3}
      \end{array} \right) \longrightarrow 
           \left( \begin{array}{ccc}
             -\cos \theta & 0 & \sin \theta \\
             0 & -1 & 0  \\
             \sin \theta & 0 & \cos \theta
           \end{array} \right)\!\!\!
                \left( \begin{array}{c} 
                  \pi_{1} \\ \pi_{2} \\ \pi_{3}
                \end{array} \right)  
\end{eqnarray}
\begin{eqnarray}
  C_{3}: \ & & (x,y,z) \longrightarrow (y,z,x) \nonumber \\
  & & \left( \begin{array}{c}
        \pi_{1} \\ \pi_{2} \\ \pi_{3}
      \end{array} \right) \longrightarrow \\
  & & \ \ \ \ \ \ \ \left( \begin{array}{ccc}
        -\sin \alpha \cos \alpha & -\cos \alpha & \sin^{2} \!\alpha \\
        -\sin \alpha & 0 & -\cos \alpha \\
        -\cos^{2} \!\alpha & -\sin \alpha \ \ & \sin \alpha \cos \alpha
      \end{array} \right)\!\!\!
           \left( \begin{array}{c} 
             \pi_{1} \\ \pi_{2} \\ \pi_{3}
           \end{array} \right) \nonumber
\end{eqnarray}
\begin{eqnarray}
  C_{5}: \ & & \left( \begin{array}{c}
                 x \\ y \\ z
               \end{array} \right) \longrightarrow 
                    \left( \begin{array}{ccc}
       \frac{1}{2} & -\frac{\sqrt{5}+1}{4} & \frac{\sqrt{5}-1}{4} \\
       \frac{\sqrt{5}+1}{4} & \frac{\sqrt{5}-1}{4} & -\frac{1}{2} \\
       \frac{\sqrt{5}-1}{4} & \frac{1}{2} & \frac{\sqrt{5}+1}{4}
                    \end{array} \right)\!\!\!
                         \left( \begin{array}{c} 
                           x \\ y \\ z
                         \end{array} \right) \nonumber \\
  & & \left( \begin{array}{c}
        \pi_{1} \\ \pi_{2} \\ \pi_{3}
      \end{array} \right) \longrightarrow
           \left( \begin{array}{ccc}
             1 & 0 & 0 \\
             0 & -\cos \frac{\pi}{5} & \sin \frac{\pi}{5} \\
             0 & -\sin \frac{\pi}{5} & -\cos \frac{\pi}{5}
           \end{array} \right)\!\!\!
                \left( \begin{array}{c} 
                  \pi_{1} \\ \pi_{2} \\ \pi_{3}
                \end{array} \right) 
\end{eqnarray}
\end{mathletters}
where $\alpha = 1.0206'$, and $\theta = 1.1410'$. Note also that
$(\pi_{1},\pi_{2},\pi_{3})$ are the fields
$(\phi^{2},\phi^{3},\phi^{4})$, in the notation used earlier in this
Section.

\section{Results.}
\label{sec:results}

\begin{figure}[b]
  \begin{center} 
    \leavevmode 
    {\hbox {\epsfxsize = 8cm \epsffile{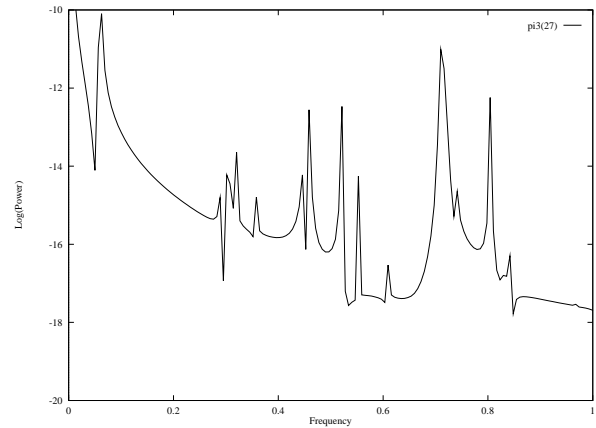}}} 
  \end{center} 
  \caption{B=7 Power Spectrum.}
    \label{fig:spectrum}
\end{figure}

Simulations of the vibrations of the B=7 multiskyrmion were carried
out on a finite lattice of $65^{3}$ points, with side length $L=8$ in
Skyrme units. The resulting power spectrum is displayed in
Figure~\ref{fig:spectrum}.

\begin{figure}[f]
  \epsfxsize = 17.5cm \epsffile{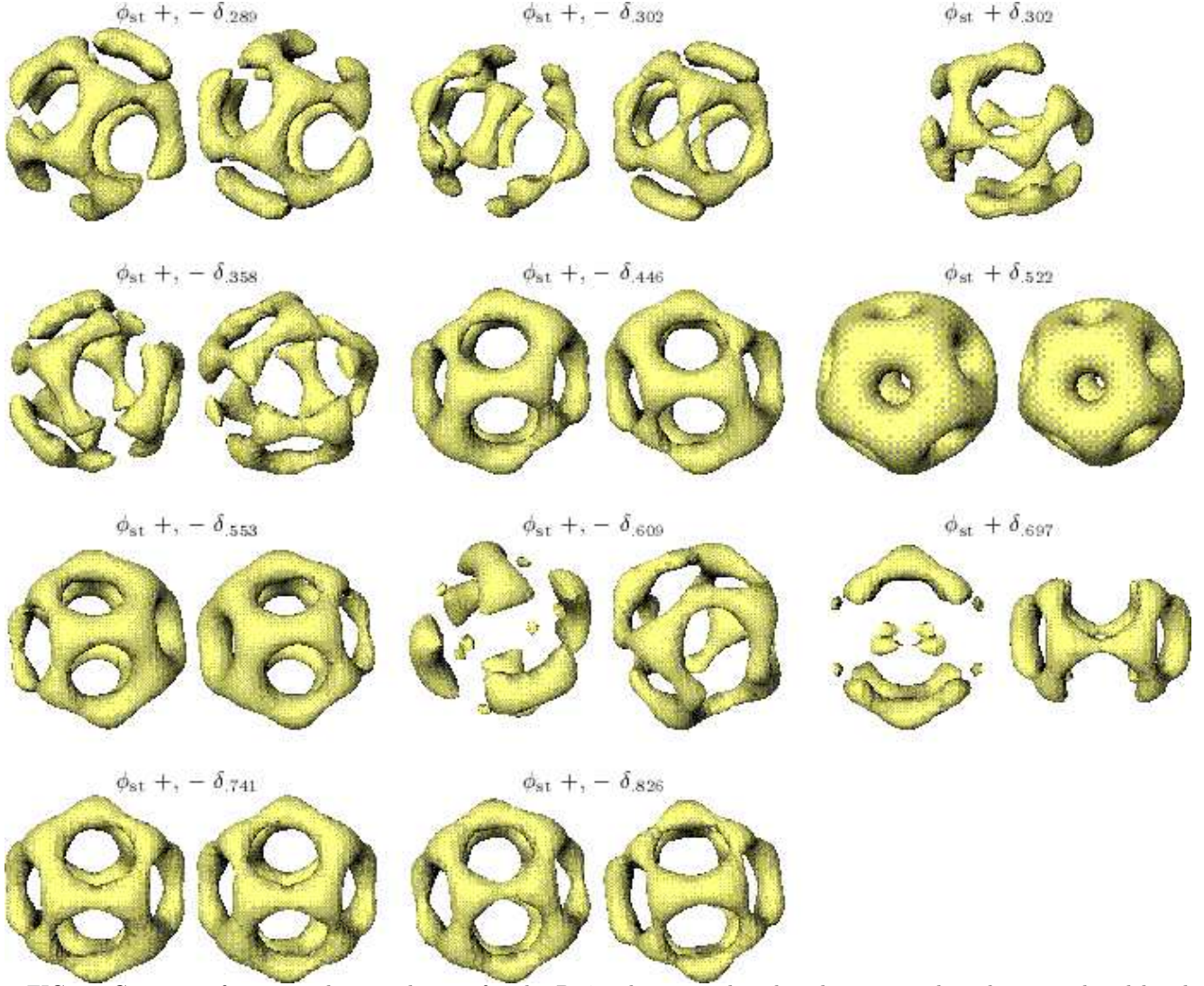}
\caption{Contours of constant baryon density for the B=7 soliton,
  combined with its normal modes, as indexed by their frequencies in
  Table~\ref{tab:modes}. Where a frequency range is given in this
  Table, modes are labelled by the lower limit of that range. Note
  that the radiation at frequency 0.609 is not plotted, so that the
  mode labelled $\delta_{.609}$ is the $4^{+}$ Singer-Sutcliffe
  scattering mode. All other modes are uniquely determined by their
  frequency or lower frequency limit, and appear in the same order as
  in Table~\ref{tab:modes}. Two different basis modes are shown for
  frequency 0.302, as the break-up into separate clusters shows up
  better in one, and the overall symmetry in the other.}
\label{fig:b7pictures}
\end{figure}

The lowest peak in the spectrum, at frequency 0.019, was found to have
total overlap $\sum_{i} \langle \delta_{.019}| J_{i}\rangle \langle
J_{i}| \delta_{.019}\rangle = .9911$ with angular momentum zero
modes. It is thus identified as comprising the rotational modes of the
static soliton, which are broken from zero to finite energy by the
discrete lattice. The same thing happened in the B=2, 3 and 4
vibrational spectra. The next three peaks contain a pure $5^{+}$
multiplet (mainly in the middle of the three peaks) and a pure $5^{-}$
multiplet, which is split between the first and third peaks, with the
modes $\{|a\rangle,|b\rangle,|c\rangle\}$ mainly in one peak, and
$\{|d\rangle,|e\rangle\}$ mainly in the other. Pictures of all the
modes may be found in Figure~\ref{fig:b7pictures}. Both of these
degeneracy 5 multiplets represent scattering type modes. In the
positive parity $5^{+}$ at frequency 0.302, two opposite pentagonal
faces of the dodecahedron split away. Presumably these will become B=2
tori, which, if the mode were amplified, would move away in opposite
directions, leaving a B=3 tetrahedron at the origin. The $5^{-}$
modes, split between frequencies 0.289 and 0.320, are more difficult
to decipher. It appears that one edge is pulled away, suggesting that
the dodecahedron may be splitting into a single skyrmion and a B=6
soliton. Both these scattering processes have a ``line'' symmetry,
which may explain their degeneracy. There are six natural lines in the
dodecahedron, the $C_{5}$ axes. Exciting modes in all six directions
simultaneously however would be equivalent to the trivial breather, so
that only five directions are really independent, hence the
degeneracy. This argument is certainly valid for the positive parity
modes. It is not as obvious for the negative parity vibrations, but
the similarity between the symmetries is certainly suggestive.

The next peak, at frequency 0.358, is a pure triplet, with symmetry
$F_{2}^{-}$. In this scattering, two pentagons from nearby, but not
adjacent, faces of the dodecahedron are pulled away. From their
relative alignments, it seems likely that these two tori will later
coalesce into a single B=4 cube. This mode probably therefore
represents the breaking of the dodecahedron into two clusters, one
with B=3 and the other B=4. A pure $4^{-}$ multiplet is split between
the next two peaks, at frequencies 0.446 and 0.459. This mode has a
tetrahedral symmetry, with four corners of the dodecahedron, arranged
in a tetrahedron, being pulled away. Presumably these will become four
single skyrmions, leaving a B=3 tetrahedron in the centre. The
opposite extreme of the motion goes through the dual tetrahedron,
hence the negative parity.

The next peak, at frequency 0.522, is a trivial breather ($1^{+}$), a
simple size fluctuation.  After that, at 0.553, is a triplet of dipole
breathing modes, with symmetry $F_{1}^{-}$. The remaining peaks all
contain a mixture of several modes. There is a pure $4^{-}$ multiplet,
spread between frequencies 0.741 to 0.842, which is a breathing mode
with tetrahedral symmetry.The last negative parity mode is a pure,
radiative $F_{2}^{-}$ triplet. This is the expected $k=0$ radiation,
which because of the damping at the beginning of the run does not
appear as a separate peak, but merely causes slight contamination
(about 20\%) in the peak at frequency 0.609.

One more reasonably pure multiplet was identified, a $4^{+}$ spread
across the frequency range 0.609 to 0.716. This is the last of the
scattering modes predicted by the monopole analogy. It breaks the
pattern previously observed in the B=2, 3 and 4 vibrational spectra,
in that it has a higher frequency than two `breathing' type
modes. This may be explained by the fact that it represents a maximal
breaking up of the B=7 soliton, splitting it into seven single
skyrmions. This mode clearly corresponds to the scattering recently
proposed by Singer and Sutcliffe \cite{SingSut}, using the instanton
approximation, as a comparison to the pictures in their paper will
demonstrate.

The remaining positive parity modes cannot be cleanly separated, but
are a mixture of one $4^{+}$ and one $5^{+}$ multiplet. In fact the
mixing is not too severe (this can be seen when computing the
characters), and the $5^{+}$ modes can be identified as quadrupole
breathing motions, while the $4^{+}$ are breathers with a line
symmetry similar to that of the $4^{+}$ scattering mode described
above. 

This completes the list of all modes found below frequency 0.9. A
summary is given in Table~\ref{tab:modes}. 
\begin{table}[t]
  \begin{center} \leavevmode 
  \begin{tabular}{|c|c|c|l|} \hline 
    Frequency & Degeneracy & Symmetry & Description \\ \hline 
     0.019 & 3 & $F_{2}^{+}$ & Broken spin zero \\
     & & & modes: $J^{+}$. \\ \hline
     0.289 - 0.320 & 5 & $5^{-}$ & Scattering: 1 + 6. \\
     & & & One edge splits away \\ 
     & & & as a single skyrmion. \\ \hline
     0.302 & 5 & $5^{+}$ & Scattering: 2 + 3 + 2. \\
     & & & Two opposite faces of \\
     & & & the dodecahedron \\
     & & & split off as B=2 tori. \\ \hline  
     0.358 & 3 & $F_{2}^{-}$ & Scattering: 3 + 4. \\
     & & & Two nearby tori split \\
     & & & away, then recombine \\
     & & & as a B=4 cube. \\ \hline
     0.446 - 0.459 & 4 & $4^{-}$ & Scattering: \\
     & & & \ \ 3 + 1 + 1 + 1 + 1. \\ 
     & & & Tetrahedral symmetry. \\ \hline
     0.522 & 1 & $1^{+}$ & Trivial breather. \\ \hline
     0.553 & 3 & $F_{1}^{-}$ & Dipole breather. \\ \hline
     0.609 & 3 & $F_{2}^{-}$ & $k=0$ radiation. \\ \hline
     0.609 - 0.716 & 4 & $4^{+}$ & Scattering: 1 + 1 + 1\\
     & & & \ \ + 1 + 1 + 1 + 1. \\
     & & & Cubic symmetry. \\ \hline
     0.697 - 0.716 & 5 & $5^{+}$ & Quadrupole breather. \\ \hline
     0.741 - 0.842 & 4 & $4^{-}$ & Breather. \\
     & & & Tetrahedral symmetry.\\ \hline
     0.826 - 0.842 & 4 & $4^{+}$ & Breather \\
     & & & Cubic symmetry. \\ \hline 
  \end{tabular} 
  \end{center}
  \caption{Summary of B=7 vibrational modes.}
  \label{tab:modes}
\end{table}
Definite frequencies are given where they can be confidently assigned;
where a mode is split into two or more peaks, or appears as part of an
inseparable mixture, a range of frequncies is given. Modes are
characterised as either scattering or breathing type, and an attempt
is made to classify the former in terms of the clusters (of different
baryon number) into which the dodecahedron breaks up. Some of these
identifications are fairly certain, while others are more tentative. A
full dynamical simulation of these attractive channels will be
necessary to confirm the guesses made here.

\section{Conclusions.}
\label{sec:conclusions}

The vibrational spectrum of the B=7 dodecahedral Skyrme soliton has
been calculated, and a full identification of the symmetries for all
modes lying below frequency 0.9 has been made. The sums defined in
Eq.~(\ref{norm}) were computed for all peaks, and all exceed 0.995, so
it seems certain that nothing has been missed at these frequencies.

All attractive channel scattering modes predicted by the monopole
analogy \cite{HMS}, were observed numerically. Four of the five
predicted complex breathing motions \cite{BM} were also observed. The
`absentee' is a predicted $5^{-}$ breathing mode. It is possible that
this multiplet may be lurking at a frequency higher than those
considered here. If so, it could be rather difficult to find it
amongst the radiation which becomes more predominant as the frequency
increases. While this mode is missing, the conjecture of Baskerville
and Michaels \cite{BM}, that there should be a total of 8B-4 modes for
a multiskyrmion of baryon number B, cannot be said to have been
conclusively proved. However, thirty eight vibrational modes have been
identified, giving a total of forty seven modes on inclusion of the
nine zero modes. This is definitely more than 6B or 6B+1 (forty two or
forty three), and while it is possible that more vibrational modes may
be discovered in the future, it is difficult to see how one could get
rid of some already found. The fact that the symmetries of the
breathing modes observed were correctly predicted by \cite{BM} is also
persuasive. Note that the instanton inspired counting suggested by
Houghton \cite{Houghton}, giving 8B-1 modes, is not ruled out
either. This ansatz gives the same vibrational modes as the
geometrical method of \cite{BM}, plus an extra triplet. If one
multiplet may have been missed due to its high frequency, perhaps a
second could be hiding up there too. The evidence would seem to favour
the mode counting scaling roughly like 8B, at any rate; whether
precisely as 8B-4 or 8B-1 remains to be decided. However, on the
whole, the predictions made for the vibrational spectrum of the B=7
Skyrme soliton have proved to be accurate.

As to the problem of the wrong spin obtained by zero mode quantisation
of the B=7 skyrmion \cite{Irwin}, there certainly exist vibrational
modes of low enough frequency to permit hope of its resolution. In
general, the more symmetry that is exhibited by a particular solution,
the greater are the constraints imposed upon its possible spin
states. Therefore, to find states with lower spin than the
$\frac{7}{2}$ permitted for the static solution, the vibrations must
break its dodecahedral symmetry. The two lowest frequency modes found
both have rather low symmetry, which is promising. Indeed, one can
only hope that these vibrations still possess enough symmetry to
forbid spin $\frac{1}{2}$ states. This calculation is beyond the scope
of this paper, but the results presented here certainly point to it as
a promising direction for future research.

\section*{Acknowledgents.}

I would like to thank N. Manton, C. Houghton, R. Battye, P. Sutcliffe
and W.J. Zakrzewski for useful discussions of this work. In
particular, P. Sutcliffe provided the computer code which generated
the approximate static solution, and N. Manton suggested some of the
scattering interpretations offered above. All simulations were
performed on COSMOS, the Origin 2000 supercomputer owned by the UK-CCC
and supported by HEFCE and PPARC. Many thanks to S. Rankin for his
help with a variety of computer-related problems. This work was
supported financially by PPARC.

\widetext
\narrowtext

\end{document}